\documentclass[a4paper,fleqn,usenatbib]{mnras}

\usepackage[T1]{fontenc}
\usepackage{ae,aecompl}

\usepackage{graphicx}	
\usepackage{amsmath}	
\usepackage{amssymb}	

\title[The Milky Way globular cluster NGC\,6809]{The surroundings of the Milky Way globular cluster NGC\,6809}

\author[Andr\'es E. Piatti]{
Andr\'es E. Piatti$^{1,2}$\thanks{E-mail: andres.piatti@unc.edu.ar} \\
$^{1}$Instituto Interdisciplinario de Ciencias B\'asicas (ICB), CONICET-UNCUYO, Padre J. Contreras 1300, M5502JMA, Mendoza, Argentina\\
$^{2}$Consejo Nacional de Investigaciones Cient\'{\i}ficas y T\'ecnicas, Godoy Cruz 2290, C1425FQB,  Buenos Aires, Argentina\\
}

\date{Accepted XXX. Received YYY; in original form ZZZ}

\pubyear{2021}

\begin{document}
\label{firstpage}
\pagerange{\pageref{firstpage}--\pageref{lastpage}}
\maketitle

\begin{abstract}
We study the outer regions of the Milky Way globular cluster NGC\,6809 based on
Dark Energy Camera (DECam) observations, which reach nearly 6 mag below the cluster 
main sequence (MS) turnoff. In order to unveil its fainter outermost structure, we built
stellar density maps using cluster MS stars, once the contamination of field
stars was removed from the cluster color-magnitude diagram. 
We found that only the resulting stellar density map  for the lightest stars 
exhibits some excesses of stars at opposite sides from the cluster  centre that diminish soon thereafter at $\sim$ 0.32$\degr$.
Studied globular clusters with apogalactic distances smaller
than that of NGC\,6809 (5.5 kpc) do not have observed tidal tails.
The lack of detection of tidal tails in the studied inner globular cluster sample
could be due to the reduced diffusion time of tidal tails by the kinematically chaotic nature
of the orbits of these globular clusters, thus shortening the time interval during which the 
tidal tails can be detected. Further investigations with an enlarged cluster sample are
needed to confirm whether chaotic and non-chaotic orbits are responsible for the
existence of globular clusters with tidal tails and those with extra-tidal features that are 
different from tidal tails or without  any signatures of extended stellar density profiles.
 \end{abstract} 




\section{Introduction}

The formation of stellar streams or tidal tails due to the stripping
or dissolution of Milky Way globular clusters has long been understood
as a consequence of their tidal interaction with their host
galaxy. Indeed, \citet{montuorietal2007} performed detailed $N$-body
simulations to show that tidal tails are generated in globular
clusters as a consequence of their interaction with the densest
components of the Milky Way (e.g. the bulge and the disc) and may result
in multi-component tidal tails after repeated apocentre
passages \citep{hb2015}. However, rather than from tidal
shocks, \citet{kupperetal2010,kupperetal2012} analytically and numerically showed
that tidal tails and the substructures within them can develop from
the epicyclic motions of a continuous stream of stars escaping the
clusters, regardless of whether the clusters' orbits are circular or
eccentric. 

From an observational point of view, there have been many
investigations of the outermost regions of globular clusters with the
goal of detecting extra-tidal structures and tidal tails. From the
early color-matched star counts of photographic plates by 
\citet{grillmairetal1995}, \citet{ls1997}, and \citet{leonetal2000}, our
sensitivity to such structures has increased by orders of magnitude
with the introduction of wide-field digital sky surveys 
\citep[e.g.,][]{odenetal2001,belokurovetal2006,gj2006,grillmair2009,gc2016,shippetal2018} and are
now reaching equivalent surface brightnesses below 35 mag/arcsec$^2$
by incorporating {\it Gaia} DR2 proper motions \citep[e.g.][]{ibataetal2019,g2019}. The results are interesting, with some globular clusters
having bona fide tidal tails, while others have irregular extended
halos or clumpy substructures, and still others have simple 
\citet{king62} radial profiles without any apparent extra-tidal features.  
\citet{pcb2020} carried out a comprehensive compilation of
the relevant observational results obtained to date with the aim of
understanding the conditions that determine whether or not a globular cluster can
develop tidal tails.  From 53 globular clusters included in
their final compilation, 14 have observed tidal tails and 17 show no
detectable signatures of extra-tidal structures.

When exploring kinematic properties (orbital eccentricity,
inclination and semi-major axis) in combination with the ratio of mass
lost by disruption to the initial cluster mass, \citet{pcb2020}
found that there are no obvious clues to differentiate
globular clusters with and without tidal tails. They also found that,
contrary to the predictions of \citet{bg2018}, globular
clusters with larger apogalactic distances and with a smaller
remaining fraction of cluster mass than Pal\,5 -a well known globular
cluster with a long tidal tail \citep{odenetal2001,gd2006} - are not necessarily 
candidates for developing tidal
tails. Furthermore, globular clusters with observed tidal tails have
apparently retained a larger fraction of their mass and have smaller
apogalactic distances than that of Pal\,5. Yet globular clusters with
extra-tidal features or \citet{king62} profiles span very similar
parameter values.  Even initial mass is uncorrelated with the presence
of tidal tails.

They also investigated whether the internal dynamical evolution of
globular clusters might be influenced by escaping stars and correlate
in some way with the presence of tidal tails. In this respect, they
considered different relationships between the core, half-mass and
Jacobi radii, the ratio of the cluster age to the respective
relaxation time and the ratio of mass loss to the total cluster
mass. The results show that, irrespective of the presence or absence
of any kind of extra-tidal structure, the globular clusters can reach
an advanced stage in their internal dynamical evolution even if they
have lost a relatively large amount of mass by tidal stripping. It
therefore seems that there is no currently known parameter that
enables us to confidently predict the presence or absence of tidal
tails for any given cluster. 

In order to enlarge the number of globular clusters from which to make meaningful 
conclusions about the existence of tidal tails, we here study NGC\,6809 (M\,55), 
an inner Milky Way globular cluster whose outermost regions has been paid little 
attention. As far as we are aware,
the cluster is included in the catalog of \citet[][2010 Edition]{harris1996}
with a tidal radius of 0.26$\degr$ \citep[see, also,][]{morenoetal2014}.
Recently, \citet{deboeretal2019} estimated a slightly larger tidal radius
(0.32$\degr$) from {\it Gaia} DR2 data. The results obtained in this work 
particularly allowed us to speculate on a possible mechanism to explain the presence 
or absence of tidal tails in globular clusters. In Sections 2 and 3 we describe the 
Dark Energy Camera 
data used and the analysis carried out  of the external cluster regions. Section 
4 discusses the resulting stellar density maps. Section 5 summarizes the main 
conclusions of  this work.

\begin{figure*}
\includegraphics[width=\textwidth]{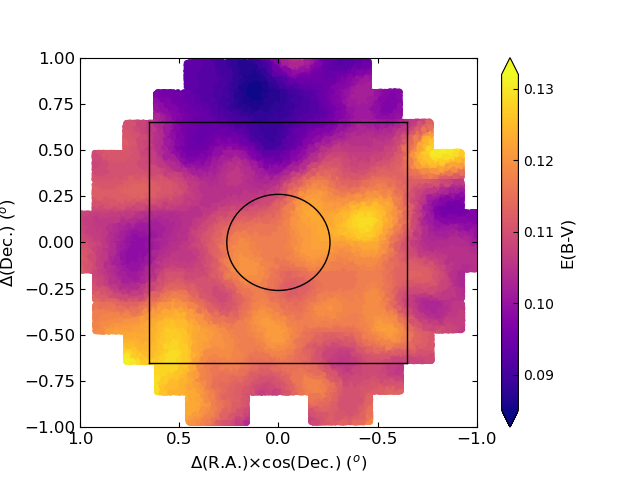}
\caption{Reddening variation across the field of NGC\,6809. The solid box 
delimites the internal boundaries of the adopted field star reference field. The circle 
corresponds to the cluster tidal radius compiled by \citet{harris1996}.}
\label{fig:fig1}
\end{figure*}

\begin{figure*}
\includegraphics[width=\textwidth]{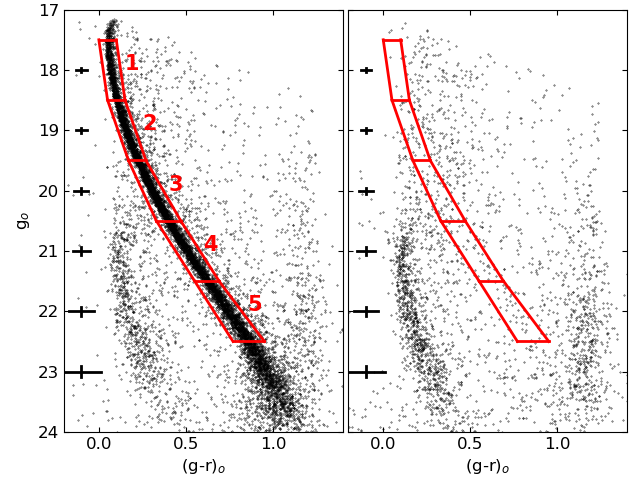}
\caption{Intrinsic CMDs for a circle of radius 
0.15$\degr$ centred on the cluster (left panel),  compared with that for an annular field region with an equal cluster area 
and internal radius of 0.8$\degr$ (right panel). Five segments along the cluster MS
are drawn and labelled in red. Typical error bars are also indicated.}
\label{fig:fig2}
\end{figure*}

\section{Data handing}

With the aim of mapping the cluster outermost structures, we used 
publicly available observations (program ID : 2019B-1003, PI : Carballo-Bello) carried out with 
the Dark Energy Camera (DECam), attached to the prime focus of the 4-m
 Blanco  telescope at Cerro Tololo Inter-American Observatory (CTIO). DECam 
 provides a 3\,deg$^{2}$ field of view with its 62 identical chips with a scale of 0.263\,arcsec\,pixel$^{-1}$ \citep{flaugheretal2015}. The gathered observations of
NGC\,6809 consists of 4$\times$600 sec $g$ and 4$\times$400 sec $r$ images,
respectively. In order to derive the  atmospheric extinction coefficients  and the transformations between the instrumental magnitudes and the SDSS $ugriz$ system \citep{fukugitaetal1996}, we used nightly observations of  5 SDSS fields at 
a different airmass.

The images were processed with the DECam Community Pipeline \citep{valdesetal2014},
while the photometry was obtained from the  images with the point-spread-function
 fitting routines of \textsc{daophot\,ii/allstar} \citep{setal90}. The final catalog
includes positions and standardized $g$ and $r$ magnitudes of stellar objects with 
$|sharpness$|$ \leq  0.5$ to avoid the presence of bad pixels, cosmic rays, galaxies, 
and unrecognized double stars in our subsequent analysis. In order to quantify the
photometry completeness, {\sc daophot\,ii}  was also employed to add synthetic stars 
with magnitudes and positions distributed similarly to those of the measured stars to an image, and carrying out the photometry for the new image as described above.  
The resulting magnitudes for the synthetic stars were then compared with those used
to create such stars. We found  that the magnitudes for a 50$\%$ completeness level
turned out to be  23.4\,mag and 23.3\,mag for the $g$ and $r$ bands, respectively 
  \citep[see, also,][]{piattietal2020,piattietal2021}.

Figure~\ref{fig:fig1} shows the variation of the interstellar reddening $E(B-V)$ across
the DECam field of view, with $E(B-V)$ values downloaded from \citet{sf11}
provided by the NASA/IPAC Infrared Science Archive\footnote{https://irsa.ipac.caltech.edu/}. 
As can be seen, $E(B-V)$ values across the entire DECam field of view span a range 
of $\sim$ 0.06 mag, which is smaller than the known lower limit for a cluster to be 
considered affected by differential reddening \citep[$\Delta$$E(B-V)$ $>$ 0.11 mag][]{burki1975}. Nevertheless,
we corrected the $g$ and $r$ magnitudes of each star by using the 
$E(B-V)$ color excesses according to the positions of the stars in the sky. 
From the dereddened $g_0$ and $r_0$ magnitudes, we built an intrinsic
color-magnitude diagram (CMD) for a circular region centred on the cluster and
with a radius of 0.15$\degr$, with the aim of highlighting the cluster features.
Figure~\ref{fig:fig2} shows the resulting CMD, as well as that for a reference field
region of equal cluster area located far from the cluster, for comparison purposes.
As can be seen, a long well populated cluster MS is clearly visible in the left panel, 
in addition to that of the Sagittarius dwarf galaxy, as first reported by \citet{mandushevetal1996}.


\section{Stellar density maps}

The strategy to build the cluster stellar density map consists in using cluster's members
distributed across the DECam field of  view. Therefore, 1) we traced five different
segments along the cluster Main Sequence (MS); then  2) we decontaminated them
of field stars, and finally  3) we built their stellar density maps with all the
measured stars that remained unsubtracted from the field star cleaning  procedure.
The five different segments were  traced to monitor any variation in the
spatial distribution of stars at large distances from the cluster centre, because
 lower-mass stars can be more easily stripped away from the cluster 
than their higher-mass counterparts.

In order to clean the cluster CMD, we followed the recipe used by \citet{pb12},
which was satisfactorily applied elsewhere  
\citep[e.g.][]{petal2018,pcb2019} 
According to that method, we need to define a cluster and a star field
areas of equal size. Figure~\ref{fig:fig1} shows a 
rectangle which delimits both regions, the latter being that outside the rectangle.
The method consists in defining boxes centred
on the magnitude and color of each star in the star field CMD; then to superimpose
them on the cluster CMD, and finally to choose the closest star to the centre of each
 box to be subtracted. 
We used CMD boxes of ($\Delta$$g_0$, $\Delta$$(g-r)_0$) =
(1.0 mag, 0.25 mag). We cleaned the cluster CMD regions delimited by the
segments traced in Fig.~\ref{fig:fig2}.



Because of the relatively large extension of the cleaned cluster area
(1.3$\degr$$\times$1.3$\degr$; see Fig.~\ref{fig:fig2}), we imposed the condition
that the spatial positions of the stars to be subtracted from the cluster MS
segment were chosen randomly. 
We then looked for a star with ($g_0$, $(g-r)_0$) values within the 
(magnitude, color) box, taking into account the
photometric errors. 
The outcome of the cleaning procedure
is a cluster MS segment that likely contains only cluster members; their 
spatial distributions relies on a random selection. For this reason, we executed 1000 
times the decontamination procedure, and defined a membership probability $P$ 
($\%$) as the ratio $N$/10, where $N$ is the number of times a star was found 
among the 1000 different outputs. In the subsequence analysis we only kept stars 
with $P$ $>$ 70$\%$.

Stellar density maps were constructed for each cleaned CMD segment 
and for stars with  fixed $P$ values
 using the \texttt{scikit-learn} software machine learning library \citep{scikit-learn} 
and its kernel density estimator (KDE). We employed a grid of 500$\times$500 
boxes on the cluster field and allowed the bandwidth to vary from 0.005$\degr$ up 
to 0.040$\degr$ in steps of 0.005$\degr$. We adopted a bandwidth of 0.025$\degr$ as
 the optimal value. The background level was estimated from the stars distributed
 in the reference star field.  We split this area in boxes 
 of 0.10$\degr$ $\times$ 0.10$\degr$ and 
counted the number of stars inside them. With the aim of enlarging the statistics,
we randomly shifted the boxes by 0.05$\degr$
along the abscisas or ordinates and repeated the star counting. 
Finally, we derived the  mean  value of the star counts coming from all the defined boxes.
We then estimated its standard deviation from a thousand Monte Carlo realisations 
of the stellar density map, shifting the positions of the stars along 
$\Delta$(RA)$\times$cos(Dec) or $\Delta$(Dec.) randomly (one different shift for 
each star) before recomputing the density map.  Figure~\ref{fig:fig3} shows the
resulting density maps that represent the deviation from the mean value in the
field in units of the standard deviation, that is,  $\eta$ = 
(signal $-$ mean value)/standard deviation,  for a signal above the mean value.  
These density maps are useful tools to identify extra-tidal features distributed not 
uniformly around the cluster's main body, as it is the case for the presence scattered 
debris, tidal tails, etc. 

\begin{figure*}
\includegraphics[width=\textwidth]{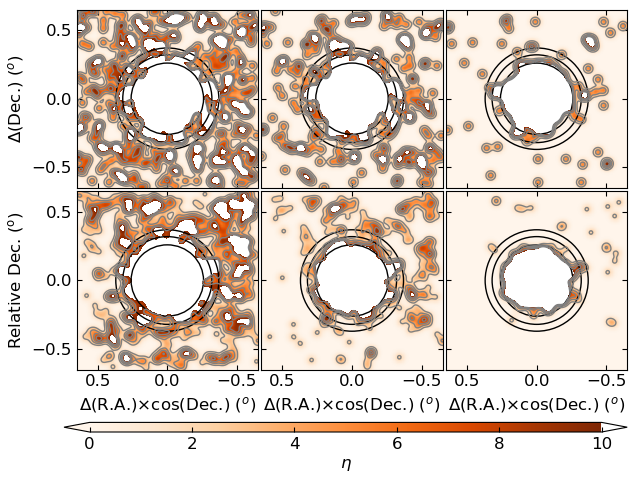}
\caption{Stellar density maps for segments 1 to 2  (from top to bottom) according 
to Fig.~\ref{fig:fig2}  and for $P> $  30, 50 and 70 $\%$ (from left to right).
The circles centred on the cluster indicate the estimated tidal radius 
0.26$\degr$ \citep{harris1996}, 0.32$\degr$ \citep{deboeretal2019}, and
the Jacobi radius for the present Galactocentric position (0.37$\degr$),
respectively. Contours for $\eta$ = 2, 4, 6, and 8 are also shown.
We have painted white stellar densities with $\eta$ $>$ 10 in order to highlight the 
least dense structures.}
\label{fig:fig3}
\end{figure*}

\setcounter{figure}{2}
\begin{figure*}
\includegraphics[width=\textwidth]{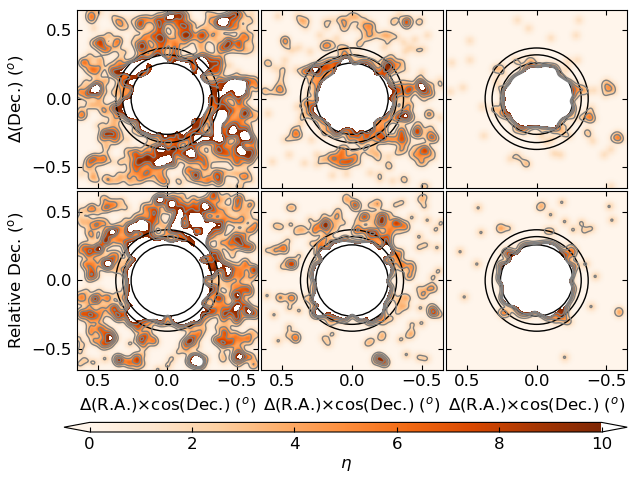}
\caption{continued, for segments 3 to 4  (from top to bottom).}
\label{fig:fig3}
\end{figure*}

\setcounter{figure}{2}
\begin{figure*}
\includegraphics[width=\textwidth]{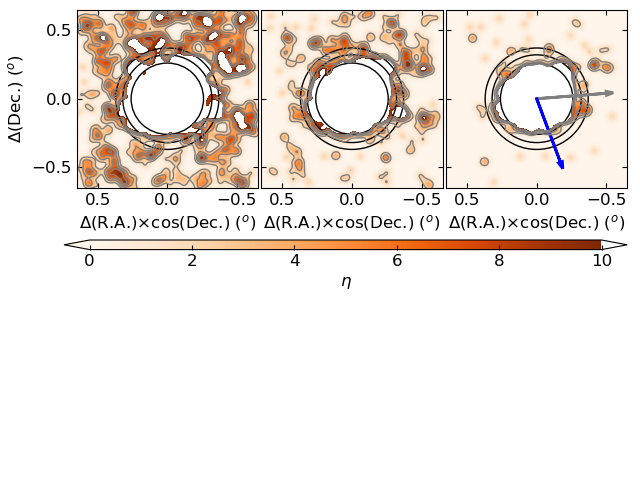}
\caption{continued,  for segment 5. The grey and blue arrows in the right panel
show the direction to the Milky Way centre and that of the motion of the cluster,
respectively.}
\label{fig:fig3}
\end{figure*}

\begin{figure*}
\includegraphics[width=\textwidth]{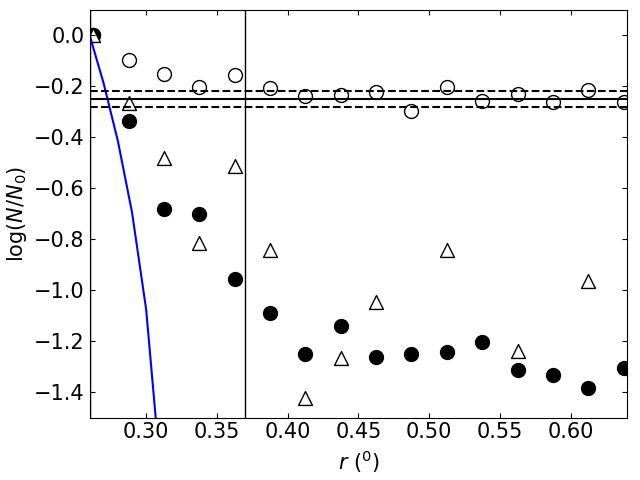}
\caption{ Normalised observed (open circle), mean background subtracted (open
triangle) stellar radial profiles and that for stars with $P$ > 70 $\%$ (filled circle).
The horizontal lines represent the mean background level and its associated
dispersion, while the vertical line represents 
the Jacobi radius for the present Galactocentric position (0.37$\degr$), respectively.
The blue solid line represents a \citet{king62}'s profile for the tidal radius 
obtained by \citet{deboeretal2019} (0.32$\degr$).} 
\label{fig:fig4}
\end{figure*}

\section{Analysis and discussion}

The produced stellar density maps of each segment look different for the
three chosen probability $P$ values. This is because a relative large area was
cleaned by spatially selecting stars to subtract randomly.
This means that the larger the $P$ values, the more similar the resulting
density maps to the intrinsic distribution of cluster stars. For this reason,
we rely our analysis on the stellar density maps built with stars with $P >$ 
70$\%$, and show those for $P>$ 30$\%$ and $>$ 50$\%$ to illustrate the 
impact of using stellar density maps with smaller $P$ values.

Stellar density maps ($P>$ 70$\%$) for the four brighter segments would not seem
to show noticeable stellar excesses beyond the tidal radius compiled by
\citet{harris1996}. This tidal radius (0.26$\degr$) readily matches the cluster extension 
(see right panels of Fig.~\ref{fig:fig3}). Note that \citet{morenoetal2014} derived the 
same tidal radius, while \citet{deboeretal2019} obtained a tidal radius slightly larger 
(0.32$\degr$). From the density map for the faintest MS stars used (segment \#5), 
excesses of stars at opposite sides from the cluster centre aligned roughly along the SE-NW 
direction for the fainter MS stars, that diminish soon thereafter,
are seen.This finding confirms that less massive 
stars are prone to leave the cluster more easily.  Nevertheless, we cannot rule out the 
presence of extended tidal tails, because some globular clusters show non- uniform tidal tails 
(see the literature compiled in \citet{pcb2020}). 

For the sake of the reader, we also built  the cluster stellar radial profile using 
all the stars distributed  in the five CMD segments (see Fig.~\ref{fig:fig2}).
We focused on the outermost 
region ($r$$>$ 0.26$\degr$) where radial variations of the photometry completeness are 
negligible and where we are interested in finding cluster extra-tidal features. 
In order to generate the stellar density profile, we counted the number of stars in annuli
of 0.025$\degr$ wide.
Figure~\ref{fig:fig4} depicts the resulting  observed radial profile represented with open 
circles. We then estimated the mean background level using all the points located at 
distances larger than 0.50$\degr$ from the cluster centre, which turned out to be 
log($N_{bg}/N_0$) = -0.25$\pm$ 0.03. Once the mean background level was subtracted from the
observed radial profile, we obtained the radial profile represented by open triangles in
Fig.~\ref{fig:fig4}. For comparison purposes we superimposed the field star cleaned
radial profile constructed from counts of stars found in the decontaminated cluster CMD
with $P$ $>$ 70$\%$ and a \citet{king62}'s model using the parameters obtained
by \citet{deboeretal2019}. As can be seen, an excess of extra-tidal stars remains in the
field star cleaned radial profile.


According to \citet{baumgardtetal2019}, NGC\,6809 describes an orbit
around the Milky Way centre characterized by an inclination angle of 
67.3$\degr$, and orbit eccentricity of 0.55, a semi-major axis of  3.6 
kpc \citep[see definition in][]{piatti2019}, and an apogalacocentric distance
of  5.6 kpc. The present cluster heliocentric and galactocentric distances are
5.3 kpc and  4.0 kpc, respectively.
Because of the cluster galactocentric distance variation, its Jacobi radius changes 
from  24.4 pc (perigalacticon) up to
39.9 pc (semi-major axis), and the cluster mass loss by tidal disruption
reaches 25$\%$ of the initial cluster mass \citep[see][]{piattietal2019b}.
The present cluster position is $\sim$ 1.5 times closer to the apogalacticon
than to the perigalacticon, and the corresponding extrapolated Jacobi radius turns out
to be  43.0 pc, which is $\sim$ 1.7 times the \citet{harris1996}'s tidal
radius, drawn in Fig.~\ref{fig:fig3}.

NGC\,6809 moves in the Milky Way with a prograde motion, which is the
direction of motion of nearly 70$\%$ of globular clusters with apogalactic
distances smaller than that of NGC\,6809 \citep{piatti2019}.  
\citet{massarietal2019} suggest that the cluster has had an accreted origin
and that the progenitor could have been a small dwarf satellite
\citep{forbes2020}. Such a cluster origin agrees well with the cluster age
\citep[13.97$\pm$0.50 Gyr; ][]{valcinetal2020}, its low metal-content
([Fe/H] = -1.99 dex; Marino et al. 2019), and the difference in
sodium abundance between first and second generation stars. Indeed,
\citet{piatti2020} found that globular clusters with  sodium
abundance enrichments larger than 0.3 dex hinted at an accreted origin,
particularly the oldest globular clusters. NGC\,6809 has a sodium
abundance enrichment of 0.72 dex \citep{piatti2020}. 


According to \citet{montuorietal2007}, the comparative denser Milky Way regions 
where NGC\,6809 is moving through should facilitate the formation of tidal tails,
which were not detected in this work. \citet{meironetal2020}, from a suite of $N$-body
 simulations, find that massive clusters should totally disrupt in the presence of
 tidal fields, while typical halo globular clusters on moderately eccentric orbits
 lose mass at a low rate that they can survive for many Hubble times. This
would not seem to match very well the case of inner Milky
 Way globulars either. Instead, the variety of prograde and retrograte orbits, with
 different eccentricities and inclinations found among globular clusters in the inner
 Milky Way rather resembles that of a kinematically chaotic
mixing system \citep[see, e.g.][]{pricewhelanetal2016a,pricewhelanetal2016b,perezvillegasetal2018}.
Recently, \citet{mestreetal2020} compared the behavior of simulated streams embedded
in chaotic and non-chaotic regions of the phase-space. They find that typical 
gravitational potentials of host galaxies can sustain chaotic orbits, which in turn
do reduce the time interval during which streams can be detected. Therefore, 
tidal tails in some globular clusters are washed out afterwards they
are generated to the point at which it is imposible to detect them. 

The percentage of mass lost in NGC\,6809 due to tidal disruption by the Milky Way 
gravitational field is within the range of values of globular clusters moving 
within a sphere of radius equals to the apogalactic
distance of NGC\,6809 \citep[between 15$\%$ and 48$\%$ of their
initial masses;][]{piatti2019,piattietal2019b}. Despite of such a relatively noticeable
amount of cluster mass lost, none out of 9 globular clusters with apogalactic distance
smaller than that of NGC\,6809 in the compilation by \citet{pcb2020} has tidal tails
identified; NGC\,5139, with an  apogalactic distance of  7.0 kpc, is
the innermost globular cluster with observed tidal tails. These findings led us to
speculate on the possibility to differentiate globular clusters with
 tidal tails from those with non-observed ones based on the chaotic nature of the
 globular clusters' orbits.  Nevertheless, it must be recalled that a non-detection 
 within 1$\degr$ does not imply the possibility of the detection on larger scales, 
 especially if supported by proper motions.
 Hence, the absence of observed tidal tails in NGC\,6809, as well as
 in other globular clusters, could be due to a comparative shorter diffusion time
 of their tidal tails because they move in a kinematically chaotic scenario.
We point out, as a caveat, the relative small number of the studied cluster sample.

\section{Conclusions}

The external regions of globular clusters, particularly their tidal tails, have become the most
sensitive tracers of the nature and distribution of dark matter in the Milky Way
\citep{bonacaetal2019}. Yet there is an ongoing debate as to whether the gaps observed in streams are due to dark matter subhalos, or to epicyclic motions of stars released from their parent clusters in discrete bursts. Particularly, \citet{diakogiannisetal2014} carried out 
a detailed dynamical analysis of NGC\,6809 and  concluded that there is no sign of 
dark matter throughout the cluster.



Here we explored the outermost regions of NGC\,6809. We built its CMD from 
DECam images centred on the cluster, which reached nearly 6 mag below the cluster MS turnoff. 
We constructed stellar density maps for stars distributed  in five different magnitude intervals along
the cluster MS. We found that only stars with a membership probability higher than 70$\%$ and $\ga$ 4 mag fainter than those at the MS  turnoff  exhibit some light
excesses of stars at opposite sides from the cluster centre aligned roughly along the SE-NW 
direction, that diminish soon thereafter, at $\sim$ 0.32$\degr$, which is the
tidal radius estimated by \citet{deboeretal2019}. The direction of the cluster
proper motion is nearly perpendicular to it, along the NE-SW vector, while the centre
of the Milky Way points to the west from NGC\,6809.

The lack of detection of  tidal tails agrees well with recent results from 
numerical simulations, which suggest that the diffusion time of streams (tidal tails
in globular clusters) is reduced by gravitational potentials that sustain
chaotic orbits, thus shortening the time interval during which the streams can be detected.
We found that
globular clusters with apogalactic distances smaller than that of NGC\,6809
have extra-tidal features that are different from tidal tails (7) or have
no signatures of extended stellar density profiles (2).
Globular clusters with detected tidal tails seem mostly to belong to the Milky Way outer halo.

\section{Data availability}

DECam images used in this work are publicly available at the https://astroarchive.noao.edu/portal/search/\#/search-form webpage.

\section*{Acknowledgements}
I thank the referee for the thorough reading of the manuscript and
timely suggestions to improve it. 

Based on observations at Cerro Tololo Inter-American Observatory, NSF’s NOIRLab (Prop. ID 2019B-1003; 
PI: Carballo-Bello), which is managed by the Association of Universities for Research in Astronomy (AURA)
under a cooperative agreement with the National Science Foundation.

This project used data obtained with the Dark Energy Camera (DECam), which was constructed by the 
Dark Energy Survey (DES) collaboration. Funding for the DES Projects has been provided by the US 
Department of Energy, the US National Science Foundation, the Ministry of Science and Education of Spain, 
the Science and Technology Facilities Council of the United Kingdom, the Higher Education Funding Council 
for England, the National centre for Supercomputing Applications at the University of Illinois at 
Urbana-Champaign, the Kavli Institute for Cosmological Physics at the University of Chicago, centre for 
Cosmology and Astro-Particle Physics at the Ohio State University, the Mitchell Institute for Fundamental 
Physics and Astronomy at Texas A\&M University, Financiadora de Estudos e Projetos, Funda\c{c}\~{a}o 
Carlos Chagas Filho de Amparo \`{a} Pesquisa do Estado do Rio de Janeiro, Conselho Nacional de 
Desenvolvimento Cient\'{\i}fico e Tecnol\'ogico and the Minist\'erio da Ci\^{e}ncia, Tecnologia e Inova\c{c}\~{a}o, the Deutsche Forschungsgemeinschaft and the Collaborating Institutions in the Dark Energy Survey.
The Collaborating Institutions are Argonne National Laboratory, the University of California at Santa Cruz, the University of Cambridge, Centro de Investigaciones En\'ergeticas, Medioambientales y Tecnol\'ogicas–Madrid, the University of Chicago, University College London, the DES-Brazil Consortium, the University of Edinburgh, the Eidgen\"{o}ssische Technische Hochschule (ETH) Z\"{u}rich, Fermi National Accelerator Laboratory, the University of Illinois at Urbana-Champaign, the Institut de Ci\`{e}ncies de l’Espai (IEEC/CSIC), the Institut de F\'{\i}sica d’Altes Energies, Lawrence Berkeley National Laboratory, the Ludwig-Maximilians Universit\"{a}t M\"{u}nchen and the associated Excellence Cluster Universe, the University of Michigan, NSF’s NOIRLab, the University of Nottingham, the Ohio State University, the OzDES Membership Consortium, the University of Pennsylvania, the University of Portsmouth, SLAC National Accelerator Laboratory, Stanford University, the University of Sussex, and Texas A\&M University.




\begin{thebibliography}{}
\makeatletter
\relax
\def\mn@urlcharsother{\let\do\@makeother \do\$\do\&\do\#\do\^\do\_\do\%\do\~}
\def\mn@doi{\begingroup\mn@urlcharsother \@ifnextchar [ {\mn@doi@}
  {\mn@doi@[]}}
\def\mn@doi@[#1]#2{\def\@tempa{#1}\ifx\@tempa\@empty \href
  {http://dx.doi.org/#2} {doi:#2}\else \href {http://dx.doi.org/#2} {#1}\fi
  \endgroup}
\def\mn@eprint#1#2{\mn@eprint@#1:#2::\@nil}
\def\mn@eprint@arXiv#1{\href {http://arxiv.org/abs/#1} {{\tt arXiv:#1}}}
\def\mn@eprint@dblp#1{\href {http://dblp.uni-trier.de/rec/bibtex/#1.xml}
  {dblp:#1}}
\def\mn@eprint@#1:#2:#3:#4\@nil{\def\@tempa {#1}\def\@tempb {#2}\def\@tempc
  {#3}\ifx \@tempc \@empty \let \@tempc \@tempb \let \@tempb \@tempa \fi \ifx
  \@tempb \@empty \def\@tempb {arXiv}\fi \@ifundefined
  {mn@eprint@\@tempb}{\@tempb:\@tempc}{\expandafter \expandafter \csname
  mn@eprint@\@tempb\endcsname \expandafter{\@tempc}}}

\bibitem[\protect\citeauthoryear{{Balbinot} \& {Gieles}}{{Balbinot} \&
  {Gieles}}{2018}]{bg2018}
{Balbinot} E.,  {Gieles} M.,  2018, \mn@doi [\mnras] {10.1093/mnras/stx2708},
  \href {http://adsabs.harvard.edu/abs/2018MNRAS.474.2479B} {474, 2479}

\bibitem[\protect\citeauthoryear{{Baumgardt}, {Hilker}, {Sollima}  \&
  {Bellini}}{{Baumgardt} et~al.}{2019}]{baumgardtetal2019}
{Baumgardt} H.,  {Hilker} M.,  {Sollima} A.,   {Bellini} A.,  2019, \mn@doi
  [\mnras] {10.1093/mnras/sty2997}, \href
  {http://adsabs.harvard.edu/abs/2019MNRAS.482.5138B} {482, 5138}

\bibitem[\protect\citeauthoryear{{Belokurov}, {Evans}, {Irwin}, {Hewett}  \&
  {Wilkinson}}{{Belokurov} et~al.}{2006}]{belokurovetal2006}
{Belokurov} V.,  {Evans} N.~W.,  {Irwin} M.~J.,  {Hewett} P.~C.,   {Wilkinson}
  M.~I.,  2006, \mn@doi [\apjl] {10.1086/500362}, \href
  {http://adsabs.harvard.edu/abs/2006ApJ...637L..29B} {637, L29}

\bibitem[\protect\citeauthoryear{{Bonaca}, {Hogg}, {Price-Whelan}  \&
  {Conroy}}{{Bonaca} et~al.}{2019}]{bonacaetal2019}
{Bonaca} A.,  {Hogg} D.~W.,  {Price-Whelan} A.~M.,   {Conroy} C.,  2019,
  \mn@doi [\apj] {10.3847/1538-4357/ab2873}, \href
  {https://ui.adsabs.harvard.edu/abs/2019ApJ...880...38B} {880, 38}

\bibitem[\protect\citeauthoryear{{Burki}}{{Burki}}{1975}]{burki1975}
{Burki} G.,  1975, \aap, \href
  {https://ui.adsabs.harvard.edu/abs/1975A&A....43...37B} {43, 37}

\bibitem[\protect\citeauthoryear{{Diakogiannis}, {Lewis}  \&
  {Ibata}}{{Diakogiannis} et~al.}{2014}]{diakogiannisetal2014}
{Diakogiannis} F.~I.,  {Lewis} G.~F.,   {Ibata} R.~A.,  2014, \mn@doi [\mnras]
  {10.1093/mnras/stt2093}, \href
  {https://ui.adsabs.harvard.edu/abs/2014MNRAS.437.3172D} {437, 3172}

\bibitem[\protect\citeauthoryear{{Flaugher} et~al.,}{{Flaugher}
  et~al.}{2015}]{flaugheretal2015}
{Flaugher} B.,  et~al., 2015, \mn@doi [\aj] {10.1088/0004-6256/150/5/150},
  \href {http://adsabs.harvard.edu/abs/2015AJ....150..150F} {150, 150}

\bibitem[\protect\citeauthoryear{{Forbes}}{{Forbes}}{2020}]{forbes2020}
{Forbes} D.~A.,  2020, \mn@doi [\mnras] {10.1093/mnras/staa245}, \href
  {https://ui.adsabs.harvard.edu/abs/2020MNRAS.493..847F} {493, 847}

\bibitem[\protect\citeauthoryear{{Fukugita}, {Ichikawa}, {Gunn}, {Doi},
  {Shimasaku}  \& {Schneider}}{{Fukugita} et~al.}{1996}]{fukugitaetal1996}
{Fukugita} M.,  {Ichikawa} T.,  {Gunn} J.~E.,  {Doi} M.,  {Shimasaku} K.,
  {Schneider} D.~P.,  1996, \mn@doi [\aj] {10.1086/117915}, \href
  {https://ui.adsabs.harvard.edu/abs/1996AJ....111.1748F} {111, 1748}

\bibitem[\protect\citeauthoryear{{Grillmair}}{{Grillmair}}{2009}]{grillmair2009}
{Grillmair} C.~J.,  2009, \mn@doi [\apj] {10.1088/0004-637X/693/2/1118}, \href
  {https://ui.adsabs.harvard.edu/abs/2009ApJ...693.1118G} {693, 1118}

\bibitem[\protect\citeauthoryear{{Grillmair}}{{Grillmair}}{2019}]{g2019}
{Grillmair} C.~J.,  2019, arXiv e-prints, \href
  {https://ui.adsabs.harvard.edu/abs/2019arXiv190905927G} {p. arXiv:1909.05927}

\bibitem[\protect\citeauthoryear{{Grillmair} \& {Carlin}}{{Grillmair} \&
  {Carlin}}{2016}]{gc2016}
{Grillmair} C.~J.,  {Carlin} J.~L.,  2016, {Stellar Streams and Clouds in the
  Galactic Halo}.
p.~87, \mn@doi{10.1007/978-3-319-19336-6_4}

\bibitem[\protect\citeauthoryear{{Grillmair} \& {Dionatos}}{{Grillmair} \&
  {Dionatos}}{2006}]{gd2006}
{Grillmair} C.~J.,  {Dionatos} O.,  2006, \mn@doi [\apjl] {10.1086/503744},
  \href {https://ui.adsabs.harvard.edu/abs/2006ApJ...641L..37G} {641, L37}

\bibitem[\protect\citeauthoryear{{Grillmair} \& {Johnson}}{{Grillmair} \&
  {Johnson}}{2006}]{gj2006}
{Grillmair} C.~J.,  {Johnson} R.,  2006, \mn@doi [\apjl] {10.1086/501439},
  \href {https://ui.adsabs.harvard.edu/abs/2006ApJ...639L..17G} {639, L17}

\bibitem[\protect\citeauthoryear{{Grillmair}, {Freeman}, {Irwin}  \&
  {Quinn}}{{Grillmair} et~al.}{1995}]{grillmairetal1995}
{Grillmair} C.~J.,  {Freeman} K.~C.,  {Irwin} M.,   {Quinn} P.~J.,  1995,
  \mn@doi [\aj] {10.1086/117470}, \href
  {http://adsabs.harvard.edu/abs/1995AJ....109.2553G} {109, 2553}

\bibitem[\protect\citeauthoryear{{Harris}}{{Harris}}{1996}]{harris1996}
{Harris} W.~E.,  1996, \mn@doi [\aj] {10.1086/118116}, \href
  {http://adsabs.harvard.edu/abs/1996AJ....112.1487H} {112, 1487}

\bibitem[\protect\citeauthoryear{{Hozumi} \& {Burkert}}{{Hozumi} \&
  {Burkert}}{2015}]{hb2015}
{Hozumi} S.,  {Burkert} A.,  2015, \mn@doi [\mnras] {10.1093/mnras/stu2318},
  \href {http://adsabs.harvard.edu/abs/2015MNRAS.446.3100H} {446, 3100}

\bibitem[\protect\citeauthoryear{{Ibata}, {Bellazzini}, {Malhan}, {Martin}  \&
  {Bianchini}}{{Ibata} et~al.}{2019}]{ibataetal2019}
{Ibata} R.~A.,  {Bellazzini} M.,  {Malhan} K.,  {Martin} N.,   {Bianchini} P.,
  2019, \mn@doi [Nature Astronomy] {10.1038/s41550-019-0751-x}, \href
  {https://ui.adsabs.harvard.edu/abs/2019NatAs...3..667I} {3, 667}

\bibitem[\protect\citeauthoryear{{King}}{{King}}{1962}]{king62}
{King} I.,  1962, \mn@doi [\aj] {10.1086/108756}, 67, 471

\bibitem[\protect\citeauthoryear{{K{\"u}pper}, {Kroupa}, {Baumgardt}  \&
  {Heggie}}{{K{\"u}pper} et~al.}{2010}]{kupperetal2010}
{K{\"u}pper} A.~H.~W.,  {Kroupa} P.,  {Baumgardt} H.,   {Heggie} D.~C.,  2010,
  \mn@doi [\mnras] {10.1111/j.1365-2966.2009.15690.x}, \href
  {http://adsabs.harvard.edu/abs/2010MNRAS.401..105K} {401, 105}

\bibitem[\protect\citeauthoryear{{K{\"u}pper}, {Lane}  \&
  {Heggie}}{{K{\"u}pper} et~al.}{2012}]{kupperetal2012}
{K{\"u}pper} A. H.~W.,  {Lane} R.~R.,   {Heggie} D.~C.,  2012, \mn@doi [\mnras]
  {10.1111/j.1365-2966.2011.20242.x}, \href
  {https://ui.adsabs.harvard.edu/abs/2012MNRAS.420.2700K} {420, 2700}

\bibitem[\protect\citeauthoryear{{Lehmann} \& {Scholz}}{{Lehmann} \&
  {Scholz}}{1997}]{ls1997}
{Lehmann} I.,  {Scholz} R.~D.,  1997, \aap, \href
  {https://ui.adsabs.harvard.edu/abs/1997A&A...320..776L} {320, 776}

\bibitem[\protect\citeauthoryear{{Leon}, {Meylan}  \& {Combes}}{{Leon}
  et~al.}{2000}]{leonetal2000}
{Leon} S.,  {Meylan} G.,   {Combes} F.,  2000, \aap, \href
  {http://adsabs.harvard.edu/abs/2000A%26A...359..907L} {359, 907}

\bibitem[\protect\citeauthoryear{{Mandushev}, {Fahlman}, {Richer}  \&
  {Thompson}}{{Mandushev} et~al.}{1996}]{mandushevetal1996}
{Mandushev} G.~I.,  {Fahlman} G.,  {Richer} H.~B.,   {Thompson} I.~B.,  1996,
  \mn@doi [\aj] {10.1086/118121}, \href
  {https://ui.adsabs.harvard.edu/abs/1996AJ....112.1536M} {112, 1536}

\bibitem[\protect\citeauthoryear{{Massari}, {Koppelman}  \& {Helmi}}{{Massari}
  et~al.}{2019}]{massarietal2019}
{Massari} D.,  {Koppelman} H.~H.,   {Helmi} A.,  2019, \mn@doi [\aap]
  {10.1051/0004-6361/201936135}, \href
  {https://ui.adsabs.harvard.edu/abs/2019A&A...630L...4M} {630, L4}

\bibitem[\protect\citeauthoryear{{Meiron}, {Webb}, {Hong}, {Berczik}, {Spurzem}
   \& {Carlberg}}{{Meiron} et~al.}{2020}]{meironetal2020}
{Meiron} Y.,  {Webb} J.~J.,  {Hong} J.,  {Berczik} P.,  {Spurzem} R.,
  {Carlberg} R.~G.,  2020, arXiv e-prints, \href
  {https://ui.adsabs.harvard.edu/abs/2020arXiv200601960M} {p. arXiv:2006.01960}

\bibitem[\protect\citeauthoryear{{Mestre}, {Llinares}  \&
  {Carpintero}}{{Mestre} et~al.}{2020}]{mestreetal2020}
{Mestre} M.,  {Llinares} C.,   {Carpintero} D.~D.,  2020, \mn@doi [\mnras]
  {10.1093/mnras/stz3505}, \href
  {https://ui.adsabs.harvard.edu/abs/2020MNRAS.492.4398M} {492, 4398}

\bibitem[\protect\citeauthoryear{{Montuori}, {Capuzzo-Dolcetta}, {Di Matteo},
  {Lepinette}  \& {Miocchi}}{{Montuori} et~al.}{2007}]{montuorietal2007}
{Montuori} M.,  {Capuzzo-Dolcetta} R.,  {Di Matteo} P.,  {Lepinette} A.,
  {Miocchi} P.,  2007, \mn@doi [\apj] {10.1086/512114}, \href
  {http://adsabs.harvard.edu/abs/2007ApJ...659.1212M} {659, 1212}

\bibitem[\protect\citeauthoryear{{Moreno}, {Pichardo}  \&
  {Vel{\'a}zquez}}{{Moreno} et~al.}{2014}]{morenoetal2014}
{Moreno} E.,  {Pichardo} B.,   {Vel{\'a}zquez} H.,  2014, \mn@doi [\apj]
  {10.1088/0004-637X/793/2/110}, \href
  {https://ui.adsabs.harvard.edu/abs/2014ApJ...793..110M} {793, 110}

\bibitem[\protect\citeauthoryear{{Odenkirchen} et~al.,}{{Odenkirchen}
  et~al.}{2001}]{odenetal2001}
{Odenkirchen} M.,  et~al., 2001, \mn@doi [\apjl] {10.1086/319095}, \href
  {http://adsabs.harvard.edu/abs/2001ApJ...548L.165O} {548, L165}

\bibitem[\protect\citeauthoryear{{Pe{\~n}arrubia}, {Varri}, {Breen}, {Ferguson}
   \& {S{\'a}nchez-Janssen}}{{Pe{\~n}arrubia}
  et~al.}{2017}]{penarrubiaetal2017}
{Pe{\~n}arrubia} J.,  {Varri} A.~L.,  {Breen} P.~G.,  {Ferguson} A. M.~N.,
  {S{\'a}nchez-Janssen} R.,  2017, \mn@doi [\mnras] {10.1093/mnrasl/slx094},
  \href {https://ui.adsabs.harvard.edu/abs/2017MNRAS.471L..31P} {471, L31}

\bibitem[\protect\citeauthoryear{Pedregosa et~al.,}{Pedregosa
  et~al.}{2011}]{scikit-learn}
Pedregosa F.,  et~al., 2011, Journal of Machine Learning Research, 12, 2825

\bibitem[\protect\citeauthoryear{{P{\'e}rez-Villegas}, {Rossi}, {Ortolani},
  {Casotto}, {Barbuy}  \& {Bica}}{{P{\'e}rez-Villegas}
  et~al.}{2018}]{perezvillegasetal2018}
{P{\'e}rez-Villegas} A.,  {Rossi} L.,  {Ortolani} S.,  {Casotto} S.,  {Barbuy}
  B.,   {Bica} E.,  2018, \mn@doi [\pasa] {10.1017/pasa.2018.16}, \href
  {http://adsabs.harvard.edu/abs/2018PASA...35...21P} {35, e021}

\bibitem[\protect\citeauthoryear{{Piatti}}{{Piatti}}{2019}]{piatti2019}
{Piatti} A.~E.,  2019, \mn@doi [\apj] {10.3847/1538-4357/ab3574}, \href
  {https://ui.adsabs.harvard.edu/abs/2019ApJ...882...98P} {882, 98}

\bibitem[\protect\citeauthoryear{{Piatti}}{{Piatti}}{2020}]{piatti2020}
{Piatti} A.~E.,  2020, \mn@doi [\aap] {10.1051/0004-6361/202039128}, \href
  {https://ui.adsabs.harvard.edu/abs/2020A&A...643A..77P} {643, A77}

\bibitem[\protect\citeauthoryear{{Piatti} \& {Bica}}{{Piatti} \&
  {Bica}}{2012}]{pb12}
{Piatti} A.~E.,  {Bica} E.,  2012, \mn@doi [\mnras]
  {10.1111/j.1365-2966.2012.21694.x}, 425, 3085

\bibitem[\protect\citeauthoryear{{Piatti} \& {Carballo-Bello}}{{Piatti} \&
  {Carballo-Bello}}{2019}]{pcb2019}
{Piatti} A.~E.,  {Carballo-Bello} J.~A.,  2019, \mn@doi [\mnras]
  {10.1093/mnras/stz500}, \href
  {https://ui.adsabs.harvard.edu/abs/2019MNRAS.485.1029P} {485, 1029}

\bibitem[\protect\citeauthoryear{{Piatti} \& {Carballo-Bello}}{{Piatti} \&
  {Carballo-Bello}}{2020}]{pcb2020}
{Piatti} A.~E.,  {Carballo-Bello} J.~A.,  2020, \mn@doi [\aap]
  {10.1051/0004-6361/202037994}, \href
  {https://ui.adsabs.harvard.edu/abs/2020A&A...637L...2P} {637, L2}

\bibitem[\protect\citeauthoryear{{Piatti} \& {Fern{\'a}ndez-Trincado}}{{Piatti}
  \& {Fern{\'a}ndez-Trincado}}{2020}]{pft2020}
{Piatti} A.~E.,  {Fern{\'a}ndez-Trincado} J.~G.,  2020, \mn@doi [\aap]
  {10.1051/0004-6361/202037439}, \href
  {https://ui.adsabs.harvard.edu/abs/2020A&A...635A..93P} {635, A93}

\bibitem[\protect\citeauthoryear{{Piatti}, {Cole}  \& {Emptage}}{{Piatti}
  et~al.}{2018}]{petal2018}
{Piatti} A.~E.,  {Cole} A.~A.,   {Emptage} B.,  2018, \mn@doi [\mnras]
  {10.1093/mnras/stx2418}, \href
  {http://adsabs.harvard.edu/abs/2018MNRAS.473..105P} {473, 105}

\bibitem[\protect\citeauthoryear{{Piatti}, {Webb}  \& {Carlberg}}{{Piatti}
  et~al.}{2019}]{piattietal2019b}
{Piatti} A.~E.,  {Webb} J.~J.,   {Carlberg} R.~G.,  2019, \mn@doi [\mnras]
  {10.1093/mnras/stz2499}, \href
  {https://ui.adsabs.harvard.edu/abs/2019MNRAS.489.4367P} {489, 4367}

\bibitem[\protect\citeauthoryear{{Piatti}, {Carballo-Bello}, {Mora}, {Cenzano},
  {Navarrete}  \& {Catelan}}{{Piatti} et~al.}{2020}]{piattietal2020}
{Piatti} A.~E.,  {Carballo-Bello} J.~A.,  {Mora} M.~D.,  {Cenzano} C.,
  {Navarrete} C.,   {Catelan} M.,  2020, \mn@doi [\aap]
  {10.1051/0004-6361/202039012}, \href
  {https://ui.adsabs.harvard.edu/abs/2020A&A...643A..15P} {643, A15}

\bibitem[\protect\citeauthoryear{{Piatti}, {Mestre}, {Carballo-Bello},
  {Carpintero}, {Navarrete}, {Mora}  \& {Cenzano}}{{Piatti}
  et~al.}{2021}]{piattietal2021}
{Piatti} A.~E.,  {Mestre} M.~F.,  {Carballo-Bello} J.~A.,  {Carpintero} D.~D.,
  {Navarrete} C.,  {Mora} M.~D.,   {Cenzano} C.,  2021, arXiv e-prints, \href
  {https://ui.adsabs.harvard.edu/abs/2021arXiv210101818P} {p. arXiv:2101.01818}

\bibitem[\protect\citeauthoryear{{Price-Whelan}, {Johnston}, {Valluri},
  {Pearson}, {K{\"u}pper}  \& {Hogg}}{{Price-Whelan}
  et~al.}{2016a}]{pricewhelanetal2016a}
{Price-Whelan} A.~M.,  {Johnston} K.~V.,  {Valluri} M.,  {Pearson} S.,
  {K{\"u}pper} A. H.~W.,   {Hogg} D.~W.,  2016a, \mn@doi [\mnras]
  {10.1093/mnras/stv2383}, \href
  {https://ui.adsabs.harvard.edu/abs/2016MNRAS.455.1079P} {455, 1079}

\bibitem[\protect\citeauthoryear{{Price-Whelan}, {Sesar}, {Johnston}  \&
  {Rix}}{{Price-Whelan} et~al.}{2016b}]{pricewhelanetal2016b}
{Price-Whelan} A.~M.,  {Sesar} B.,  {Johnston} K.~V.,   {Rix} H.-W.,  2016b,
  \mn@doi [\apj] {10.3847/0004-637X/824/2/104}, \href
  {https://ui.adsabs.harvard.edu/abs/2016ApJ...824..104P} {824, 104}

\bibitem[\protect\citeauthoryear{{Schlafly} \& {Finkbeiner}}{{Schlafly} \&
  {Finkbeiner}}{2011}]{sf11}
{Schlafly} E.~F.,  {Finkbeiner} D.~P.,  2011, \mn@doi [\apj]
  {10.1088/0004-637X/737/2/103}, 737, 103

\bibitem[\protect\citeauthoryear{{Shipp} et~al.,}{{Shipp}
  et~al.}{2018}]{shippetal2018}
{Shipp} N.,  et~al., 2018, \mn@doi [\apj] {10.3847/1538-4357/aacdab}, \href
  {https://ui.adsabs.harvard.edu/abs/2018ApJ...862..114S} {862, 114}

\bibitem[\protect\citeauthoryear{{Stetson}, {Davis}  \& {Crabtree}}{{Stetson}
  et~al.}{1990}]{setal90}
{Stetson} P.~B.,  {Davis} L.~E.,   {Crabtree} D.~R.,  1990, in {Jacoby} G.~H.,
  ed.,  Astronomical Society of the Pacific Conference Series Vol. 8, CCDs in
  astronomy. pp 289--304

\bibitem[\protect\citeauthoryear{{Valcin}, {Bernal}, {Jimenez}, {Verde}  \&
  {Wandelt}}{{Valcin} et~al.}{2020}]{valcinetal2020}
{Valcin} D.,  {Bernal} J.~L.,  {Jimenez} R.,  {Verde} L.,   {Wandelt} B.~D.,
  2020, \mn@doi [\jcap] {10.1088/1475-7516/2020/12/002}, \href
  {https://ui.adsabs.harvard.edu/abs/2020JCAP...12..002V} {2020, 002}

\bibitem[\protect\citeauthoryear{{Valdes}, {Gruendl}  \& {DES
  Project}}{{Valdes} et~al.}{2014}]{valdesetal2014}
{Valdes} F.,  {Gruendl} R.,   {DES Project} 2014, in {Manset} N.,  {Forshay}
  P.,  eds,  Astronomical Society of the Pacific Conference Series Vol. 485,
  Astronomical Data Analysis Software and Systems XXIII. p.~379

\bibitem[\protect\citeauthoryear{{de Boer}, {Gieles}, {Balbinot},
  {H{\'e}nault-Brunet}, {Sollima}, {Watkins}  \& {Claydon}}{{de Boer}
  et~al.}{2019}]{deboeretal2019}
{de Boer} T.~J.~L.,  {Gieles} M.,  {Balbinot} E.,  {H{\'e}nault-Brunet} V.,
  {Sollima} A.,  {Watkins} L.~L.,   {Claydon} I.,  2019, \mn@doi [\mnras]
  {10.1093/mnras/stz651}, \href
  {https://ui.adsabs.harvard.edu/abs/2019MNRAS.485.4906D} {485, 4906}

\makeatother
\end{thebibliography}







\bsp	
\label{lastpage}
\end{document}